\documentclass[aps,11pt,showpacs,showkeys]{revtex4}
\usepackage{amsmath}
\usepackage{graphicx}
\usepackage{natbib}
\usepackage{color}
\begin{document}
\title{Advances in atomtronics}
\begin{abstract}
Atomtronics is a relatively new subfield of atomic physics that aims to realize the device behavior of electronic components in ultracold atom-optical systems.  The fact that these systems are coherent makes them particularly interesting since, in addition to current, one can impart quantum states onto the current carriers themselves or perhaps perform quantum computational operations on them.  After reviewing the fundamental ideas of this subfield, we report on the theoretical and experimental progress made towards developing externally-driven and closed loop devices.  The functionality and potential applications for these atom analogs to electronic and spintronic systems is also discussed.
\end{abstract}
\author{R. A. Pepino}
\email{rpepino@flsouthern.edu}

\affiliation{Florida Southern College, Lakeland, FL 33801}
\pacs{01.30.Rr, 03.65.Yz, 03.67.-a,03.75.Kk, 03.75.Lm, 05.70.Ln, 06.20.-f}
\keywords{atomtronics, ultracold gas, quantum simulation, quantum computing}
\maketitle

\section{Introduction}
Quantum simulation has become a major research effort in atomic physics over the last three decades.  Bose-Einstein condensates [BECs] have been generated, trapped and manipulated in countless experiments and the introduction of optical lattices has allowed for the experimental realization of condensed matter phenomena such as the Mott insular to superfluid phase transition and the Hofstadter butterfly~\citep{greiner2002quantum,PhysRevLett.111.185301}.  As optical techniques have improved it has become possible to produce, not only honeycomb~\citep{lee2009ultracold} and Kagom\'e lattices~\citep{jo2012ultracold}, but custom optical lattices using holographic masks~\citep{zupancic2016ultra}.  Customized potential well structures for ultracold atoms can also be generated using atom chips~\citep{tosto2019optically},  accoustic optical modulators [AOMs]~\citep{henderson2009experimental} and digital micromirror devices [DMDs]~\citep{gauthier2016direct}.

In this article, we review a relatively new research trajectory in quantum simulations with ultracold atoms known as {\em atomtronics}.  One of the goals of this subfield is to realize behavior analogous to traditional electronic devices with neutral, ultracold atomic systems~\citep{Brian}.  Atom analogs for batteries, diodes, transistors, logic gates and superfluid quantum interference devices [SQUIDs] have been proposed~\citep{pepino2009atomtronic,pepino2010open,stickney2007transistorlike,zozulya2013principles,wright2013driving}.  Some of these porposals have already been realized in experiments~\citep{caliga2016transport,caliga2017experimental,krinner2017two}.  There are several reasons to investigate these types of systems: the transport characteristics in novel systems can be studied~\citep{krinner2015observation}, the inherent matter-wave characteristics of these systems could lead to high precision measurement applications~\citep{grun2020integrable}, in a computational context, the coherent current dynamics could allow for additional information to be imprinted on the current carrier's internal quantum states and finally, these platforms might also find themselves integrated into quantum computing architectures~\citep{sinuco2016addressed}.

This article is organized into two main sections.  The first section discusses the dynamics of atomtronic devices that are coupled to external reservoirs which act as  sources and drains for neutral, ultracold atomic current. The second section considers devices made out of BECs that are trapped in ring configurations.  In these systems, atomic current is excited by external laser fields.  We review theoretically-proposed and experimentally-realized devices.  We also discuss some of their potential applications.  

\section{Externally-Driven Devices}
In this section, we discuss atomtronic schemes that are directly analogous to their electronic counterparts in the sense that they are being driven by an external power source.  The systems we focus on here are composed of optical lattices whose system characteristics can be modeled by the Bose-Hubbard Hamiltonian~\citep{zoller}
\begin{equation}
\hat{H}_{BH}=\sum_{j=1}^N\epsilon_j\hat{a}_j^\dagger\hat{a}_j+
\frac{1}{2}U_j\hat{a}_j^\dagger\hat{a}_j^\dagger\hat{a}_j\hat{a}_j+
\sum_{\langle j,k\rangle}J_{jk}\hat{a}_j^\dagger\hat{a}_{k}+h.c.
\label{bhh}
\end{equation}
where $j$ and $k$ are  lattice site indices, $\epsilon_j$ and $U_j$ are the external and on-site interaction energies for the $j$th lattice cite, and $J_{jk}$ is the tunneling rate between sites $j$ and $k$.

\begin{figure}[h]
\includegraphics[width=10.5 cm]{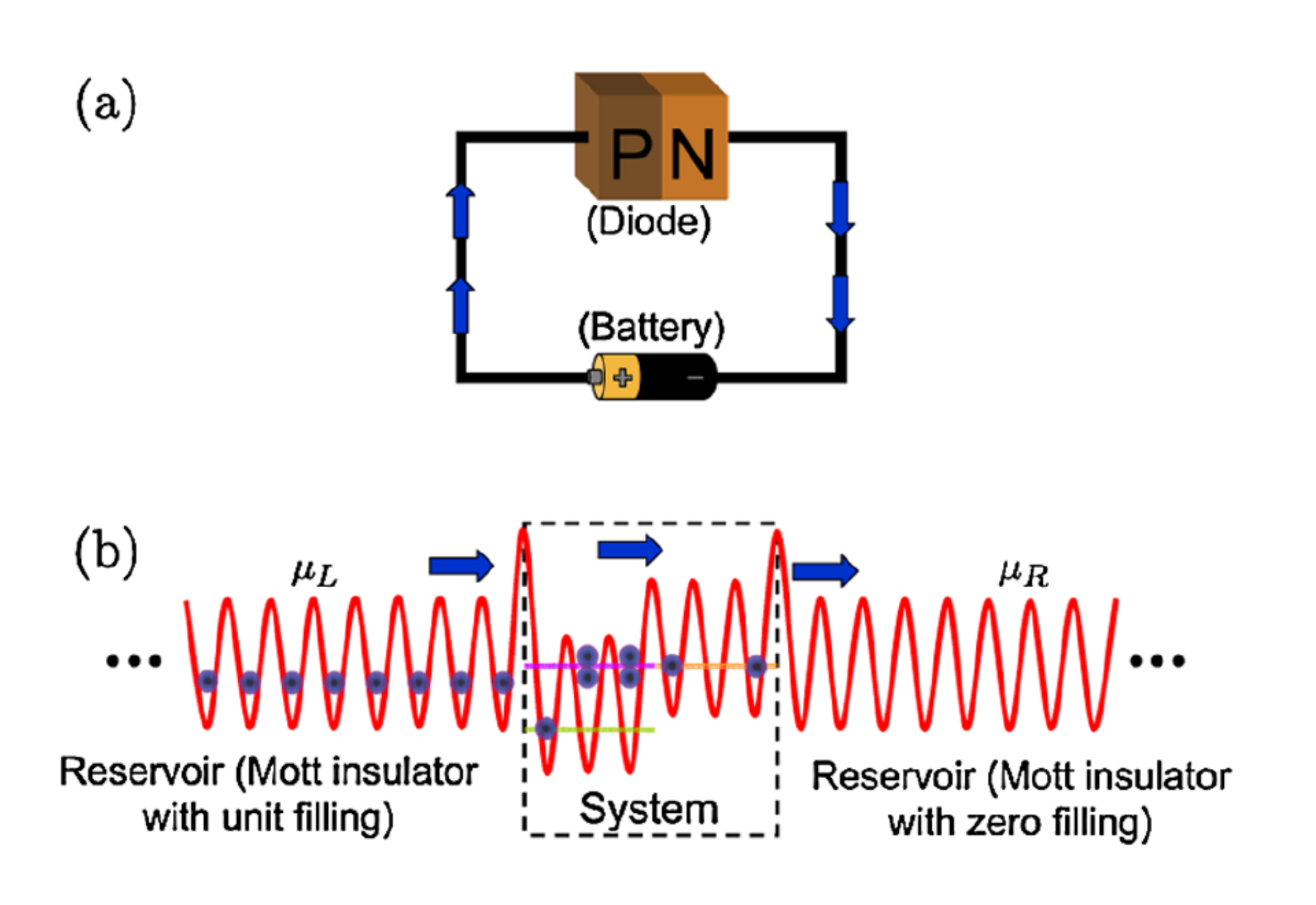}
\caption{Figure taken from Reference~\cite{pepino2009atomtronic}.  ({\bf a}) An illustration of a traditional electronic p-n junction diode connected to an external power source.  ({\bf b})  The equivalent atomtronic circuit where an optical lattice with an external energy gap labeled `System' is attached to two external optical lattice reservoirs, which drive neutral, ultracold atoms across the system by acting as a source (left) and sink (right).  The external energy gap of the system is responsible for generating the diode-like response in the optical lattice}
\label{circuit}
\end{figure}
In place of the electric potential difference of a battery, atomtronic components are driven by a chemical potential difference.  As depicted in Figure~\ref{circuit}(b), reservoirs of ultracold atoms are connected to opposite ends of the system.  When a chemical potential difference exists between the reservoirs, the reservoir with the higher chemical potential acts as a source of atoms while the lower acts as a drain.

\subsection{Atomtronic batteries}
One proposal for a battery is a coherent drive consisting of a BEC trapped in a potential well~\citep{zozulya2013principles}. If such a drive was connected to the first lattice site, the system Hamiltonian would be
\begin{equation}
\hat{H}=\Omega \left(\hat{a}_1+\hat{a}_1^\dagger\right)+H_{BH}
\end{equation}
where $\Omega$ is the system-BEC tunneling rate.  A potential issue of the coherent drive battery is that its dynamics are inherently reversible.  This may ultimately compromise the effectiveness of the battery as a current source: the neutral atomic current has the same chance of entering the system as it does returning to the battery.  This becomes a potentially greater issue when considering the dynamics of the drain side of the battery. In order to remove the reversibility in the system, the experimental realization of the BEC drive required utilizing a `terminator beam', which consisted of a resonant laser that removed atoms once they entered the drain trap~\citep{caliga2016transport}.   In a different attempt to combat transport issues with coherent sources, it has been proposed to dynamically alter the depths of the source/drain wells.  However, simulations of these dynamics did not yield very promising results~\citep{lai2016challenges}.

An alternative realization of a battery would be to couple the ends of the system to large Markovian-like reservoirs of ultracold atoms.  The irreversibility of the system-reservoir interaction ensures a one way flow of neutral atomic current.  In most regimes, one could employ the standard Born and Markov approximations and end up modeling the system-reservoir coupling as a Lindblad master equation~\citep{cohen1998atom}.  However, it was found that divergences arose when the chemical potentials of the reservoirs overlapped with the energies of the system.  In order to tame these divergences, an all-orders ansatz for the memory kernel was added to the system~\citep{pepino2010open}.

Such reservoirs could be constructed out of large 2- or 3-D optical lattices themselves.  However a parameter hierarchy would need to be honored in order for the reservoir to correctly function as a source or a sink.  Specifically, the tunneling rates of the reservoir must be much faster than those of the system, and both of these tunneling rates have to be fast compared to that of the atoms entering or leaving the system.  The last condition, which assures a weak system-reservoir coupling, guarantees that the memory associated with the system-reservoir interaction has enough time to reset between particle exchanges.  This hierarchy, validates the mathematical assumptions made in Reference~\citep{pepino2010open}.

\subsection{Diodes, transistors and logic gates}
Diodes and transistors are electronic components with distinct device characteristics.  A diode only allows current to flow in one direction, while a transistor exhibits switching and/or amplification of a signal.  It has been shown that diode-like and transistor-like behavior can be realized by creating certain energetic resonances within the atom-lattice system~\citep{pepino2009atomtronic}.  Constructing the necessary resonances requires site-by-site tunability of the energies of the optical lattice.  Using holographic masks, AOMs and/or DMDs, it  is possible to generate custom optical lattices, or other multi-well systems, with the required level of precision.

Diode-like behavior is exhibited in a flat optical lattice with a specific step function-like external energy as depicted in Figure~\ref{circuit}(b). Since the dynamics do not depend on the number of lattice sites, we explain the functionality of the atomtronic diode consisting of only two sites here.

\begin{figure}[h]
\includegraphics[width=10.5 cm]{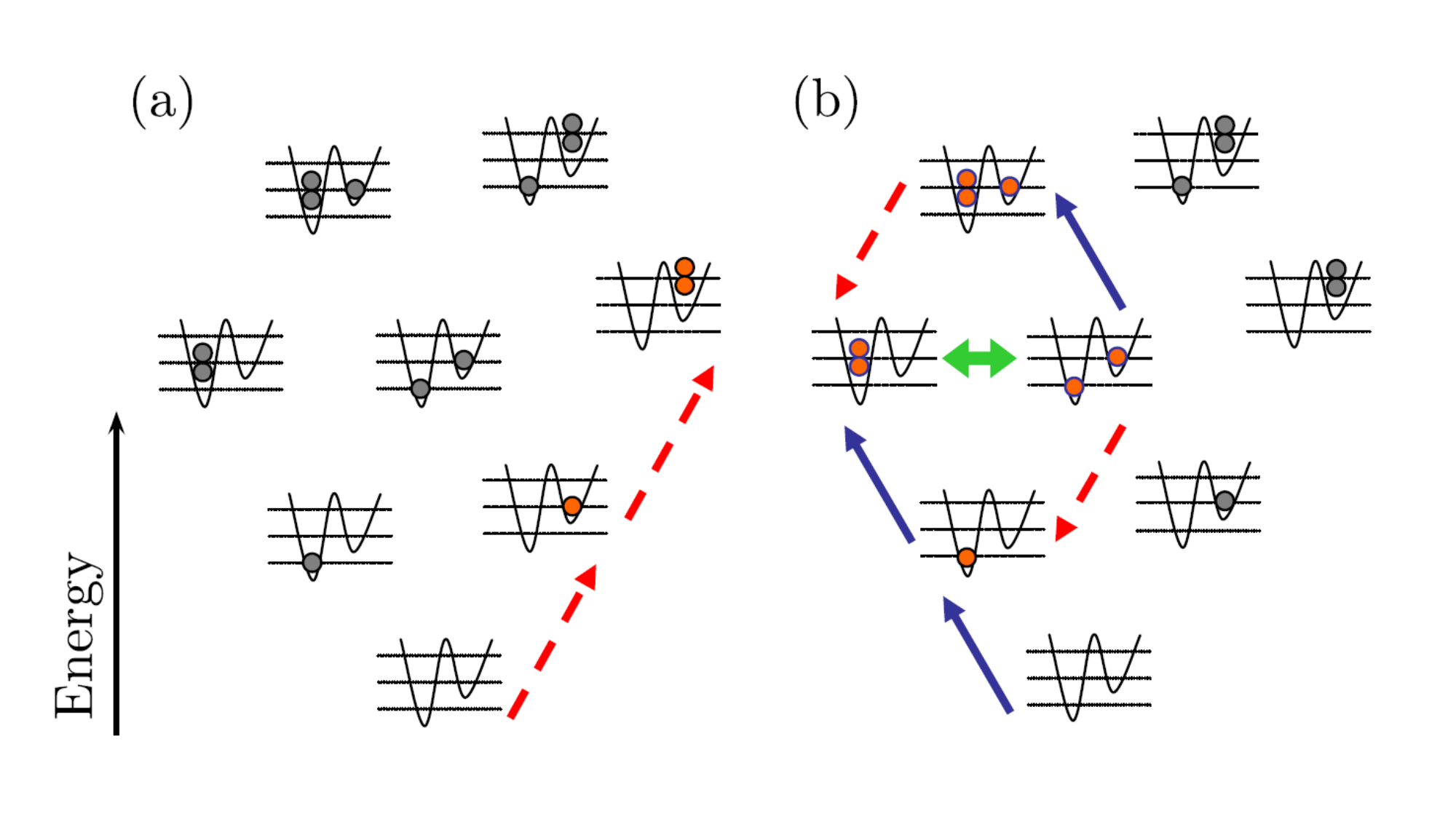}
 \caption{
Figure taken from Reference~\citep{pepino2010open}.  An illustration of the system's Fock states energies.  The blue solid and red dashed arrows represent the action of the left and right reservoirs respectively.  ({\bf a}) The energy schematic of the reverse bias dynamics of the two site atomtronic diode. Regardless of which state the system starts in, it evolves almost entirely to the $|02\rangle$ state. ({\bf b}) The energy schematic of the forward-bias dynamics of the two-site atomtronic diode. The resonance between the $|20\rangle$ and $|11\rangle$ states, in combination with the action of the reservoir's attempt to keep a population two atoms on the left site while leaving the right site vacant, leads to the following two cycles of states that lead to current flow through the system: $|20\rangle\to|11\rangle\to|10\rangle\to|20\rangle$ and $|20\rangle\to|11\rangle\to|21\rangle\to|20\rangle$.}
\label{resonances}
\end{figure}

If the external energy of the upper lattice site is equal to the external energy of the lower site plus the on-site interaction energy, $\epsilon_2=\epsilon_1+U$, there is a resonance between the $|20\rangle$ and $|11\rangle$ system Fock states.  This resonance is depicted in  the Fock state energy schematics of Figure~\ref{resonances}.  As seen in Figure~\ref{resonances}(a), if the reservoirs are situated to maintain an occupancy of two atoms on the second lattice site and remove all from the first, the isolation of the $|01\rangle$ and $|02\rangle$ states ensures that virtually no atomic current traverses through the system.  On the other hand, if the first lattice site was being driven, once the $|20\rangle$ state is accessed, its resonance with $|11\rangle$ enables atoms to traverse to the second site of the lattice.  The reservoir action of trying to keep two atoms on the left and zero on the right initiates the two intrasystem cycles, depicted in Figure~\ref{resonances}(b), which lead to a current flowing through the system.

Bipolar junction transistor-like behavior can be achieved with three reservoirs (functioning as the collector, base and emitter components of a traditional transistor) connected to as little as three lattice sites.  In this case, if $\epsilon_1=\epsilon_3=\epsilon_2+U$, then a resonance is generated between the three states $|110\rangle$, $|020\rangle$ and $|011\rangle$~\citep{pepino2009atomtronic}.  Assuming that the chemical potentials of left and right reservoirs are set to maintain an occupancy of one atom on the left site and zero on the right, the chemical potential of the reservoir connected the middle site determines whether or not current flows through the system.  If it is set to maintain an occupancy of zero atoms, the system finds itself predomanently in the $|100\rangle$ state.  However, if the middle reservoir is set to maintain an occupancy of one atom on middle site, the resonance between the $|110\rangle$, $|020\rangle$ and $|011\rangle$ Fock states is accessed and current is able to flow through the system.  The gain between the base and emitter sites is determined by the relative coupling strengths of the middle reservoir compared to the other two.  A closed system version of this transistor was also proposed in Reference~\citep{stickney2007transistorlike} and has since been realized experimentally~\cite{caliga2016transport}.  It was also shown that two of these types of transistors coupled in series or parallel reproduce AND and OR logic gate characteristics in the same way their classical counterparts do~\citep{pepino2010open}.  In order to generate more complicated components for either transport or computational purposes, polychromatic amplitude modulation may be needed~\citep{pepino2016transport}.

Transistor-like behavior was proposed in other systems as well.  Once such transistor consists of three lattice sites where either a BEC or Fermi gas is prepared on the left and a fermion impurity is fixed in the middle~\citep{micheli2004single}.  This transistor scheme functions in a way analogous to electromagnetically-induced transparency: if the impurity atom is prepared in one state, the impurity is transparent to the current carriers.  If the impurity is prepared in its other state, quantum interference forbids atomic current from going through the system.  Another three well transistor scheme was proposed that involves two species repulsively-interacting bosons~\citep{gajdacz2014atomtronics}.  This transistor has a feature that allows it to function as a CNOT gate.  A photovoltaic transistor emulator has also been proposed~\citep{lai2019photovoltaic}. Finally, an atomtronic analog of the spintronic Datta-Das transistor has been proposed~\citep{vaishnav2008spin}, and experimentally realized~\citep{beeler2013spin}, in ultracold atomic systems.  Since the spintronic Datta-Das transistor itself has not yet been realized experimentally, the atomtronic experiment was the first one to realize its physics.

\section{Closed Loop Circuits}
An alternative approach to atomtronics involves closed loop circuits of BECs trapped in toroidal or ring lattice potentials.  The superfluid current dynamics of these isolated quantum systems have coherent properties that resemble superconducting quantum interference devices (SQUIDs).  This opens up the possibility for atomtronic systems to be utilized as ultra-sensitive rotation sensors as well as magnetometers.  It has also been shown that flux qubits can be created within these systems. The long coherence times of these particular systems makes them promising candidates for quantum computing platforms.

\subsection{Atomtronic SQUIDs and quantum sensing}
\begin{figure}[h]
\includegraphics[width=10.5 cm]{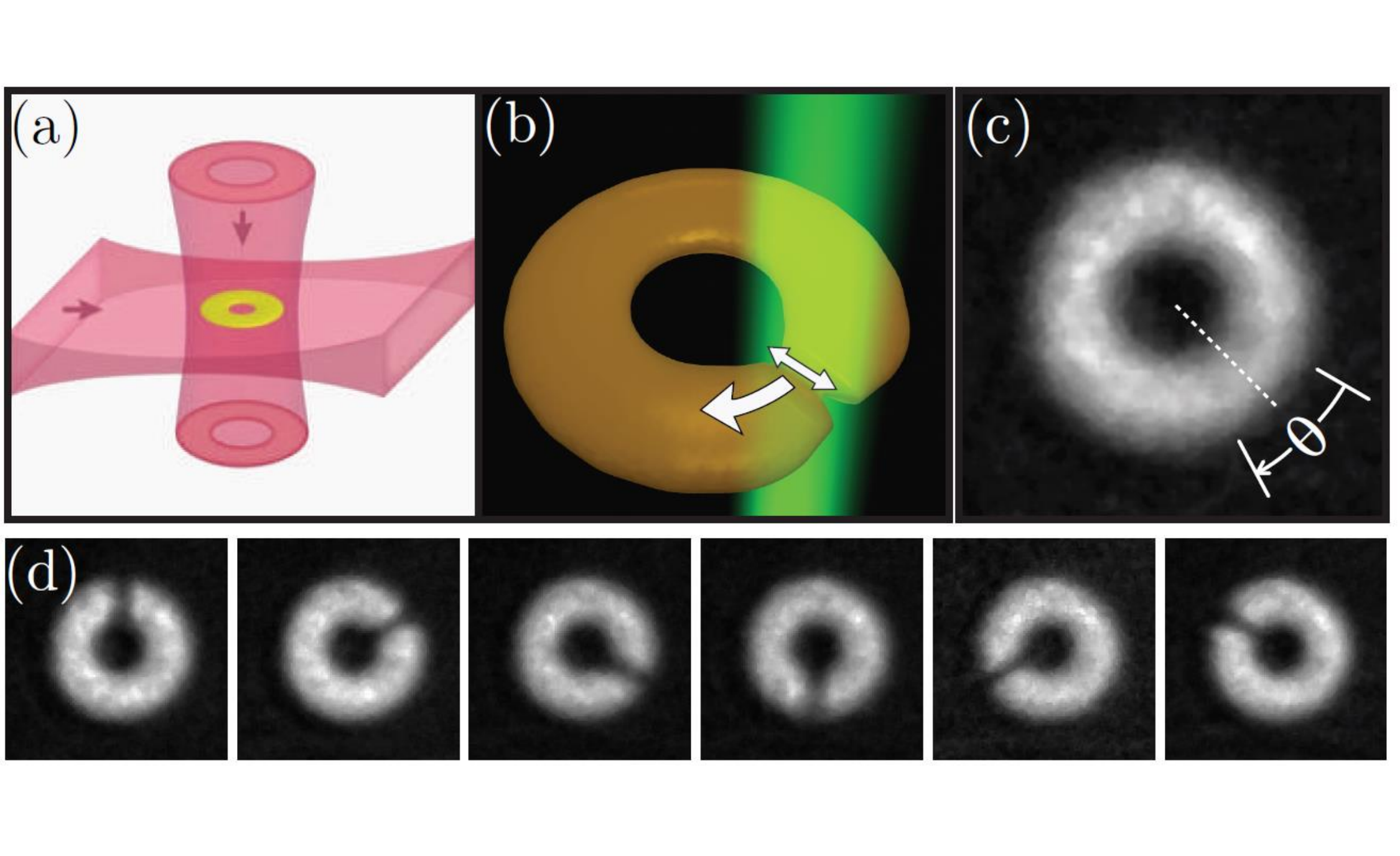}
\caption{Figure taken from Reference~\citep{wright2013driving}. ({\bf a}) A schematic of the optical dipole trap, ({\bf b}) the geometry of the barrier beam, ({\bf c}) experimental absorption image of the ring condensate with $\theta$ marking the arc of the barrier trajectory and ({\bf d}) experimental images of the barrier's angular positions at integer multiples of $60^\circ$.}
\label{GretchenFig}
\end{figure}

About ten years ago, a series of experiments involving BECs confined to toroidal potentials that were stirred with different numbers of rotating lasers started taking place.  In addition to rotating the BECs in a circular motion, the lasers generate low superfluid density regions.  In the laser's reference frame, the system is equivalent to a superconducting current loop in the presence of an external magnetic field with internal weak links.  The first experiment, depicted in Figure~\ref{GretchenFig}, was an atomtronic realization of a single-junction DC SQUID~\citep{wright2013driving}.  The role that SQUIDs play in magnetometers and gyroscopes make these systems relevant for metrological applications~\citep{clarke2004squid}.  The second generation of these experiments  introduced an additional moving barrier, which realized circuitry similar to that of a standard DC Squid~\citep{ryu2013experimental,jendrzejewski2014resistive,ryu2020quantum}.  It was later proposed that currents of entangled {\em solitons} could exhibit SQUID-like behavior in similar single barrier experiments involving bosons with an attractive interaction~\citep{polo2020quantum}.

Recently, there have been proposals to generate N00N states in closed ring lattice systems.  One of these proposals consists of a driven ring lattice  whose sites have an external energy gap~\citep{compagno2020multimode}, while another involves a four site square lattice containing bosonic atoms with a dipolar interaction~\citep{grun2020integrable}.  The maximally-entangled nature of N00N states might make these systems potentially useful in quantum measurements as well.

\subsection{Quantum computation with atomtronics}
The long coherence times of these isolated, neutral, ultracold atom-optical systems potentially make them promising platforms for quantum computing.  There have been  proposals for realizing flux qubits in the types of closed loop circuits discussed above~\citep{gallemi2015coherent,aghamalyan2016atomtronic}.  The optics necessary for constructing scalable architectures for these types of qubits in stacked and planar arrays have already been experimentally realized~\citep{safaei2018two,amico2014superfluid}.  A potential benefit these quantum computing systems have over those involving single atom qubits is their relatively large physical spacing, which would make addressing single qubits easier.

Hysteresis has also been observed in these loop circuit systems~\citep{eckel2014hysteresis}.  The existence of hysteresis opens up the door for  atomtronic systems to act as computational memory elements, and possibly  digital noise filters as well.

\section{Conclusions}
In this article, we have surveyed some of the recent progress made in developing ultracold atom-optical analogs of electronic components.  The two trajectories in the field of atomtronics currently involve externally-driven devices, and closed loop current systems.  By exploiting intrasystem resonances, one can tune optical lattices (and other potential well structures) to mimic the behavior of fundamental electronic components such as batteries, diodes, both conventional and Datta-Das transistors, as well as basic logic gates.  In the loop circuit context, neutral BEC systems can be constructed to mimic the behavior of a variety of electronic SQUIDs.  The transport control capabilities of the externally-driven devices, along with the long coherence times these systems might also lead to them playing some role in future quantum computing schemes.  Experiments have shown that the technology for creating and manipulating atomtronic circuits and devices currently exists.  Future work in this field might include utilizing these systems to make ultra sensitive rotation/magnetic sensors, integrating existing components to form more complicated quantum circuitry, encoding quantum states onto atomic currents, and possibly developing new components altogether.

\bibliographystyle{unsrt}
\bibliography{mybib}

\end{document}